\documentclass[twocolumn,prb,amsmath,amssymb,floatfix,superscriptaddress]{revtex4-1}
\usepackage{graphicx}
\usepackage{dcolumn}
\usepackage{bm}
\usepackage{xspace}
\usepackage{hyperref}
\usepackage{color}

\providecommand*{\tc}{$T_c$\xspace}%
\begin{document}

\title{Unusual phonon density of states and response to superconducting transition in the In-doped topological crystalline insulator Pb$_{0.5}$Sn$_{0.5}$Te}
\author{Kejing~Ran}
\affiliation{National Laboratory of Solid State Microstructures and Department of Physics, Nanjing University, Nanjing 210093, China}
\author{Ruidan~Zhong}
\altaffiliation{Present address: Department of Chemistry, Princeton University, Princeton, New Jersey 08544, USA}
\affiliation{Condensed Matter Physics and Materials Science
Department, Brookhaven National Laboratory (BNL), Upton, New York 11973,
USA}
\affiliation{Materials Science and Engineering Department, Stony Brook University, Stony Brook, New York 11794, USA}
\author{Tong~Chen}
\author{Yuan~Gan}
\author{Jinghui~Wang}
\affiliation{National Laboratory of Solid State Microstructures and Department of Physics, Nanjing University, Nanjing 210093, China}
\author{B.~L.~Winn}
\author{A.~D.~Christianson}
\affiliation{Neutron Scattering Division, Oak Ridge National Laboratory (ORNL), Oak Ridge, Tennessee 37831, USA.}
\author{Shichao~Li}
\author{Zhen~Ma}
\author{Song~Bao}
\author{Zhengwei~Cai}
\affiliation{National Laboratory of Solid State Microstructures and Department of Physics, Nanjing University, Nanjing 210093, China}
\author{Guangyong~Xu}
\altaffiliation{Present address: NIST Center for Neutron Research, National Institute of Standards and Technology, Gaithersburg, Maryland 20899, USA}
\affiliation{Condensed Matter Physics and Materials Science
Department, Brookhaven National Laboratory, Upton, New York 11973, USA}
\author{J.~M.~Tranquada}
\author{Genda~Gu}
\affiliation{Condensed Matter Physics and Materials Science
Department, Brookhaven National Laboratory, Upton, New York 11973, USA}
\author{Jian~Sun}
\email{jiansun@nju.edu.cn}
\author{Jinsheng~Wen}
\email{jwen@nju.edu.cn}
\affiliation{National Laboratory of Solid State Microstructures and Department of Physics, Nanjing University, Nanjing 210093, China}
\affiliation{Collaborative Innovation Center of Advanced Microstructures, Nanjing University, Nanjing 210093, China}


\begin{abstract}
We present inelastic neutron scattering results of phonons in (Pb$_{0.5}$Sn$_{0.5}$)$_{1-x}$In$_x$Te powders, with $x=0$ and 0.3. The $x=0$ sample is a topological crystalline insulator, and the $x=0.3$ sample is a superconductor with a bulk superconducting transition temperature $T_c$ of 4.7~K.
In both samples, we observe unexpected van Hove singularities in the phonon density of states at energies of 1--2.5~meV, suggestive of local modes.
On cooling the superconducting sample through $T_c$, there is an enhancement of these features for energies below twice the superconducting-gap energy. We further note that the superconductivity in (Pb$_{0.5}$Sn$_{0.5}$)$_{1-x}$In$_x$Te occurs in samples with normal-state resistivities of order 10 m$\Omega$~cm, indicative of bad-metal behavior.
Calculations based on density functional theory suggest that the superconductivity is easily explainable in terms of electron-phonon coupling; however, they completely miss the low-frequency modes and do not explain the large resistivity. While the bulk superconducting state of (Pb$_{0.5}$Sn$_{0.5}$)$_{0.7}$In$_{0.3}$Te appears to be driven by phonons, a proper understanding will require ideas beyond simple BCS theory.

\end{abstract}

\maketitle


Topological insulators (TIs) represent an exotic state of matter in which the bulk is insulating but the surface is metallic \cite{hasa10,qi11}. The topological state is protected by the time-reversal symmetry~\cite{hasa10}. By including a certain crystal point group symmetry instead of time reversal, topological crystalline insulators (TCIs), a state analogous to TIs, was also proposed~\cite{ando15,fu11}. In particular, it was predicted that compounds such as SnTe might be TCIs \cite{hsie12}, and the key features, including the inverted character of bands near the chemical potential \cite{xu12,tana12} and surface states within the band gap that are protected from back-scattering \cite{assa14,akiy14}, were soon verified.

These developments have also spurred renewed interest in topological superconductors, as such materials may exhibit gapless surface states that could be beneficial for quantum computing \cite{hasa10,qi11,ando15}. A common aspect of TCIs is strong spin-orbit coupling in the atomic states contributing to the valence and conduction bands, and this is a useful ingredient for obtaining the unusual pairing symmetry expected for a topological superconductor.  Hence, there is interest in seeing whether doping TCIs can induce superconductivity with a topological nature.   Indeed, one can make SnTe superconducting by introducing In, and a point-contact study found evidence of surface Andreev bound states \cite{sasa12}, though thermodynamic studies suggest a fully gapped superconducting state \cite{he13,sagh14,smyl18}.

SnTe is a compound that has the rocksalt structure despite the fact that the component elements do not all form closed-shell ions.  A consequence is a strong intrinsic electron-phonon coupling \cite{luco73,litt79}, which leads to a ferroelectric phase transition in SnTe \cite{koba76,onei17}.  One way to suppress the ferroelectric transition is to substitute Pb for Sn, and it has been estimated that the transition should hit 0~K for Pb$_{y}$Sn$_{1-y}$Te with $y\approx 0.5$ \cite{daug78}. At this Pb concentration the system becomes a TCI \cite{tana13}. Also, substitution of a small amount of In turns the system into a true bulk insulator \cite{zhon15}, while adding more In makes it superconducting \cite{parf01,zhon14}. For reference, we note that the proximity of the superconductivity and the suppressed ferroelectric order suggests a possible connection with the enhanced superconductivity at the ferroelectric quantum critical point in SrTiO$_3$ \cite{risc17,rowl14}.

\begin{figure*}[t]
\centering
\includegraphics[width=0.95\linewidth]{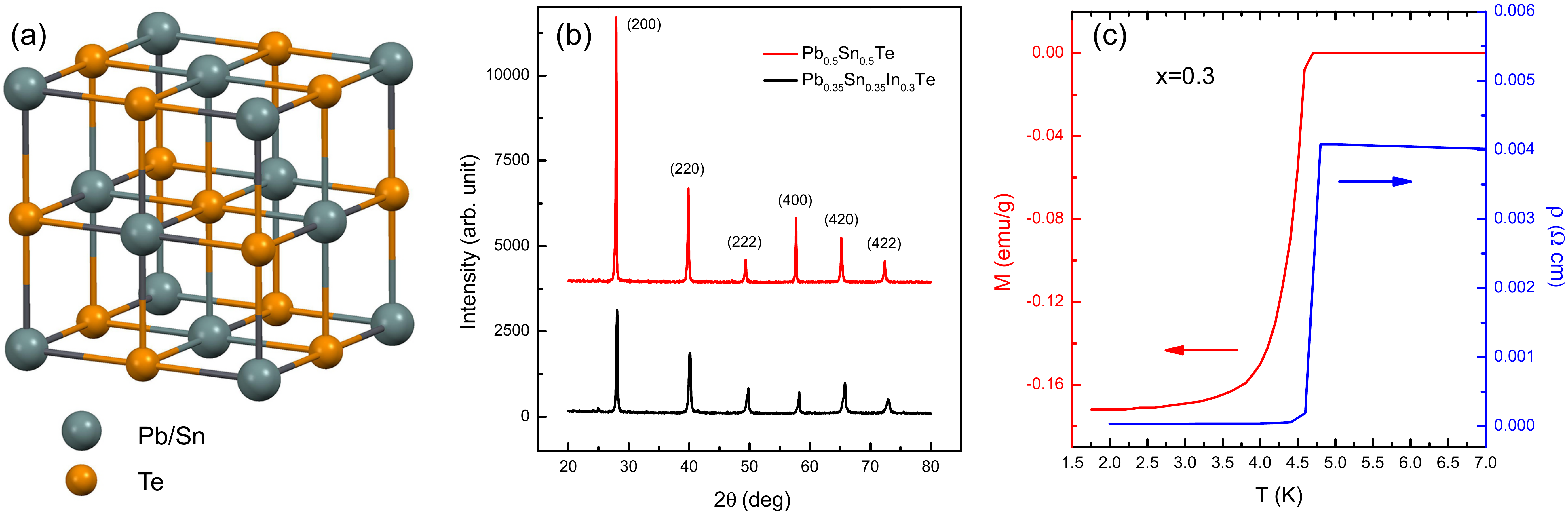}
\caption{\label{fig:structure}{(a) Schematic for the crystal structure of Pb$_{0.5}$Sn$_{0.5}$Te. (b) X-ray powder diffraction results for (Pb$_{0.5}$Sn$_{0.5}$)$_{1-x}$In$_x$Te with $x=0$ and 0.3 measured at room temperature. (c) Temperature dependence of the magnetization (left axis) and resistivity (right axis) for (Pb$_{0.5}$Sn$_{0.5}$)$_{0.7}$In$_{0.3}$Te single crystals. The magnetizations were measured in zero-field-cooling conditions with a magnetic field of 10~Oe.}}
\end{figure*}

In this Rapid Communication, we use inelastic neutron scattering (INS) to measure the phonon density of states (PDOS) in polycrystalline samples of (Pb$_{0.5}$Sn$_{0.5}$)$_{1-x}$In$_x$Te (PSIT) with $x=0$ (a nonsuperconducting TCI), and $x=0.3$ (a superconductor with a critical temperature $T_c$ of 4.7~K). In both samples, we detect unexpected peaks, in the range of 1--2.5 meV, that are not reproduced by a calculation using the virtual crystal approximation. While calculations of the electron-phonon coupling indicate contributions from phonon features from 3 to 12~meV that should easily be sufficient to explain the $T_c$, we observe a surprising enhancement below $T_c$ of phonon intensities with energies less than twice that of the $T=0$ superconducting gap.  We also note that the normal-state resistivity is unusually large and nonmetallic, possibly associated with the electron-phonon coupling; in any case, it raises questions regarding the existence of well-defined quasiparticles and the applicability of conventional BCS theory \cite{bcstheory}.

Single crystals of PSIT with $x=0$ and 0.3 were grown by the modified floating-zone technique as described in Refs~\cite{zhon14,PhysRevB.88.020505,crystals7_55}. The sample compositions were measured with a scanning electron microscope (SEM) equipped with an analyzer for energy dispersive x-ray spectroscopy. About ten positions were measured for each crystal piece, and the variation in $x$ was generally found to be $<2\%$ of the mean value. Moreover, the SEM images showed that there is no secondary phase for the samples studied here\cite{zhon14}. Fine powders were obtained by grinding these crystals inside a glovebox filled with Ar. Powder x-ray diffraction (XRD) measurements were performed at room temperature on a Rigaku Miniflex II located at Center for Functional Nanomaterials at BNL, using Cu $K\alpha$ radiation. The diffraction patterns, plotted in Fig.~\ref{fig:structure}(b), can be nicely indexed with the rocksalt structure (space group $Fm\bar{3}m$) shown in Fig.~\ref{fig:structure}(a). These data indicate that impurity level is below 1\%. Room-temperature lattice constants are determined to be 6.392 and 6.359~{\AA} for the $x=0$, and 0.3 samples, respectively.

Magnetization and resistivity were measured on single-crystal pieces extracted from the same batches as those used to make the powders for XRD and INS measurements. Magnetization and resistivity were measured using a Quantum Design magnetic properties measurement system and physical properties measurement system, respectively. The sample with no In is a TCI~\cite{tana13,zhon14,crystals7_55}. It shows a weakly metallic behavior and is not superconducting down to 1.8~K. The $x=0.3$ sample is superconducting with a $T_c$ of 4.7~K, as determined from the onset of diamagnetism and zero resistivity, as shown in Fig.~\ref{fig:structure}(c). This is the optimal $T_c$ of this system, before the solubility of $x=0.35$ is reached~\cite{zhon14,crystals7_55}.
The $x=0.3$ sample shows a sharp superconducting transition in both the magnetization and resistivity measurements, indicating that the sample is homogeneous, consistent with the XRD results and composition analysis.

\begin{figure}[b]
\centering
\includegraphics[width=0.98\linewidth]{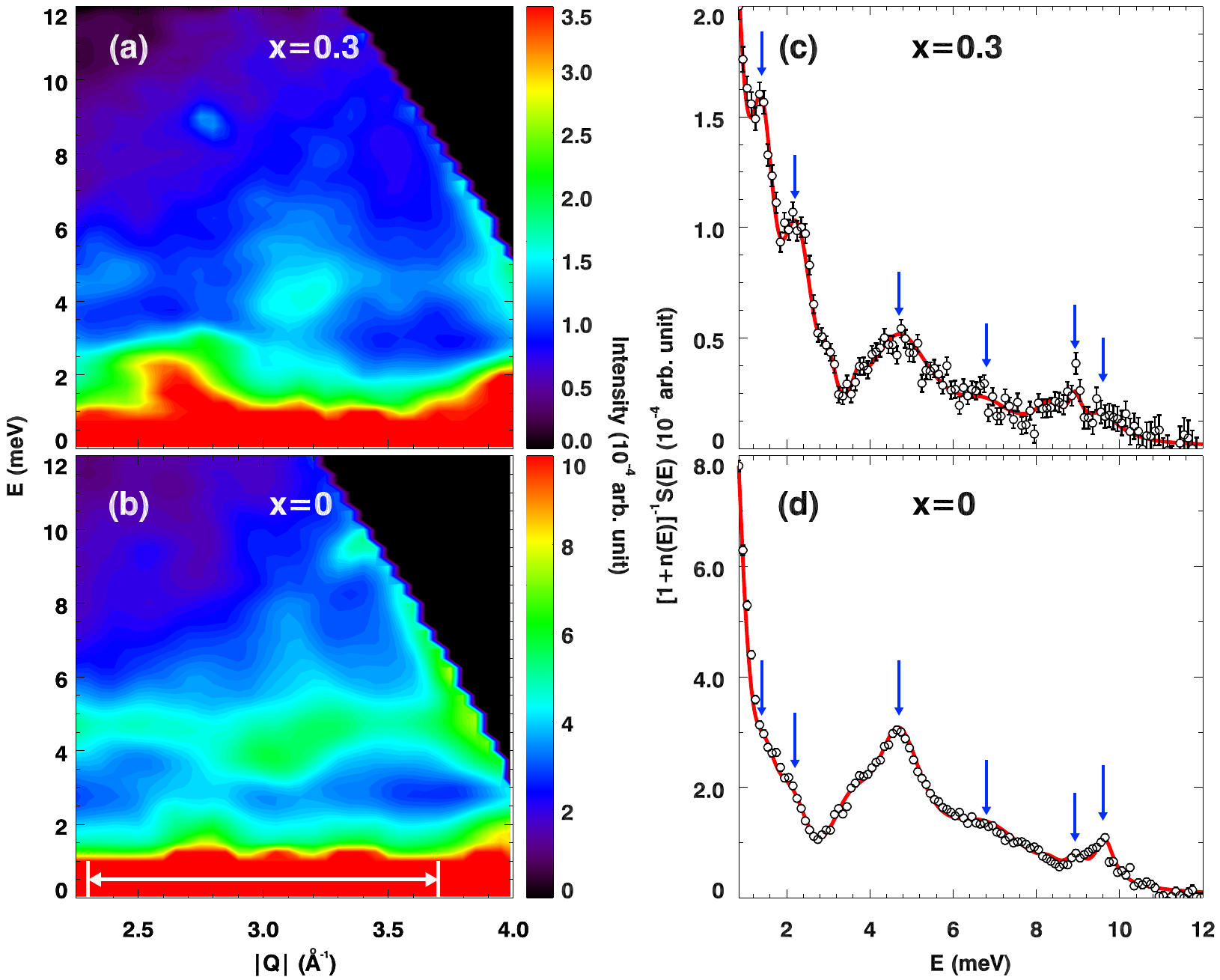}
\caption{\label{fig:map}{(a) and (b) Inelastic neutron scattering results for (Pb$_{0.5}$Sn$_{0.5}$)$_{1-x}$In$_x$Te with $x=0.3$ and 0, respectively, measured at 7~K. (c) Scattering intensities (corrected by the Bose factor) as a function of energy obtained by integrating the intensities in (a) with $|\bm{Q}|$ ranging from 2.2 to 3.7~\AA$^{-1}$ [indicated in (b)] for the $x=0.3$ sample. (d) Same as (c), but for the $x=0$ sample. Lines through data are guides to the eye. Vertical arrows represent major phonon modes. Error bars represent one standard deviation throughout this Rapid Communication.}}
\end{figure}

We used 20-g powders well characterized as discussed above for each compound in the INS experiments, carried out on two time-of-flight (TOF) spectrometers HYSPEC~\cite{epj83_03017} and ARCS, both located at the Spallation Neutron Source (SNS) at ORNL. On HYSPEC and ARCS, we chose an incident energy $E_{i}$ of 15 and 35~meV, respectively. On HYSPEC, we used a Fermi frequency of 360~Hz, which gave an energy resolution of $\sim$0.4~meV (full width at half maximum) at the elastic position. Only the low-energy data collected on HYSPEC are presented in this work.
The results at 7~K (above the $T_c$ of the superconducting sample) are shown in Figs.~\ref{fig:map}(a) and ~\ref{fig:map}(b). Since In absorbs neutrons strongly~\cite{neutron1}, the scattering intensities for the In-doped sample are much weaker than those of the In-free sample. The scattering around 5~meV is prominent, with the strength increasing with increasing $|\bm{Q}|$, where $\bm{Q}$ is the wave vector. Such a $\bm{Q}$ dependence of the scattered intensity is consistent with those resulting from phonons. We integrate the intensities $S(\bm{Q},E)$ over $\bm{Q}$ ranging from 2.2 to 3.7~{\AA}, as indicated in Fig.~\ref{fig:map}(b), divide them by the detailed-balance factor $[1+n(E)]$, where $n(E)$ is the Bose distribution function, and plot them as a function of energy in Figs.~\ref{fig:map}(c) and ~\ref{fig:map}(d). Prominent features are indicated by the arrows in Figs.~\ref{fig:map}(c) and ~\ref{fig:map}(d). Although the positions for both samples appear to be the same, the relative intensities among different features show some differences. These indicate that although the In doping does not significantly alter the crystal structure, it does affect the lattice dynamics.

\begin{figure}[b]
\centering
\includegraphics[width=0.95\linewidth]{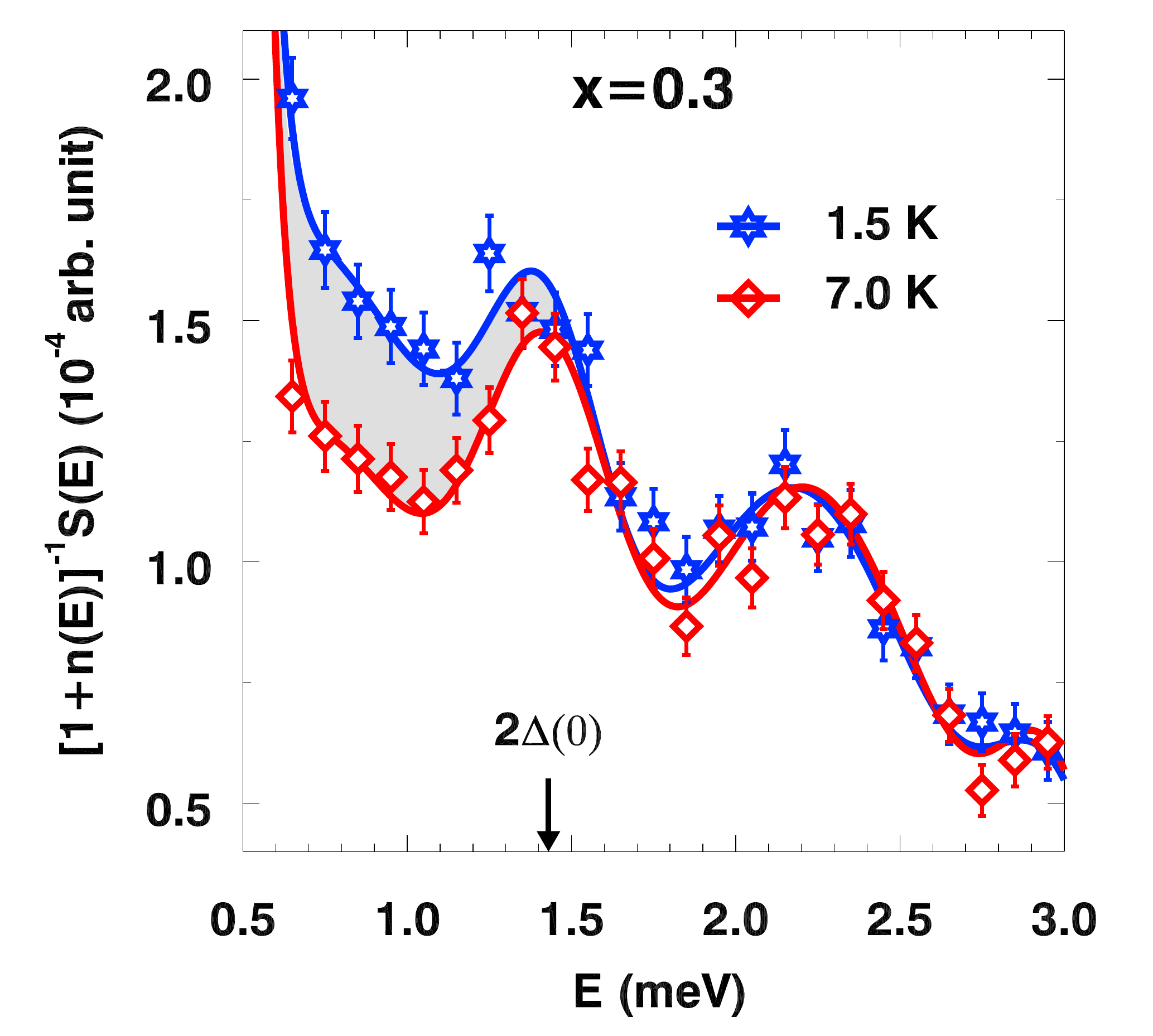}
\caption{\label{fig:pdos}{Low-energy phonon spectra for (Pb$_{0.5}$Sn$_{0.5}$)$_{0.7}$In$_{0.3}$Te at 1.5 and 7~K. Intensities $S(E)$ obtained by integrating the intensities in Fig.~\ref{fig:map}(a) with $|\bm{Q}|$ ranging from 2.5 to 3.0~\AA$^{-1}$ have been corrected by the Bose factor. The arrow indicates twice of the estimated zero-temperature superconducting gap, $2\Delta(0)$. Lines through data are guides to the eye. The shadow illustrates the difference between the low- and high-temperature data.}}
\end{figure}

The measurements on the $x=0.3$ sample were repeated at 1.5~K, well below $T_c$. The only significant changes across $T_c$ occur at low energies, as illustrated in Fig.~\ref{fig:pdos}. At 1.5~K, the intensities are enhanced compared to those at 7~K for energies below $\sim$1.4~meV. Interestingly, this energy is almost exactly twice of the zero-temperature superconducting gap, 2$\Delta(0)=1.44$~meV, obtained in the same material from scanning tunneling spectroscopy measurements~\cite{PhysRevB.92.020512}. The ratio of 2$\Delta(0)$ over $k_{\rm{B}}T_c$ is about 3.5, consistent with the BCS prediction, corresponding to the weak-coupling regime~\cite{bcstheory}. This ratio is also consistent with that obtained in a similar system, Sn$_{1-x}$In$_x$Te~\cite{sagh14}. For reference, the non-superconducting Pb$_{0.5}$Sn$_{0.5}$Te sample has also been measured at 1.5~K, but there was no significant change relative to 7~K. The enhancement of the phonon scattering only in the superconducting sample below $T_c$, with a characteristic energy equal to 2$\Delta(0)$, indicates an intriguing and highly unusual correlation between the low-energy phonons and the superconducting order.

\begin{figure*}[hpt]
\centering
\includegraphics[width=0.98\linewidth]{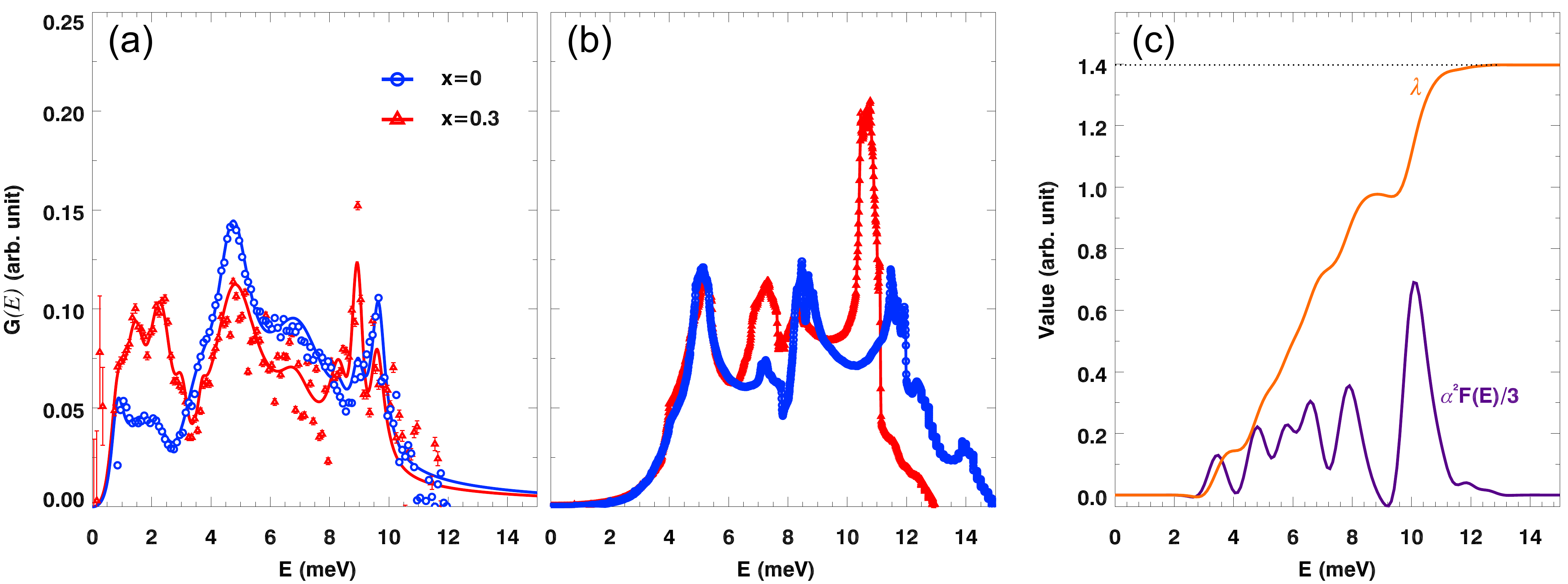}
\caption{\label{fig:pdos2}{(a) Experimental results of the weighted PDOS for (Pb$_{0.5}$Sn$_{0.5}$)$_{1-x}$In$_{x}$Te with $x=0$ and 0.3, measured at 7~K. Lines through data are guides to the eye. (b) Calculated results of the PDOS for the two samples. (c) The calculated Eliashberg spectral function $\alpha^2F(E)$ for  (Pb$_{0.5}$Sn$_{0.5}$)$_{0.7}$In$_{0.3}$Te, scaled by 1/3, and the total electron-phonon coupling constant $\lambda$. The dashed line indicates the $\lambda$ value of 1.4.}}
\end{figure*}

To gain further insight, we have converted the $|\bm{Q}|$-integrated scattered intensities $S(E)$ into the PDOS $G(E)=\sum_{i}\frac{\sigma_{i}}{m_{i}}{\rm e}^{-2W_{i}}G_{i}(E)$, where $\sigma_i$, $m_i$, $W_i$, and $G_i$ are the neutron scattering cross section, atomic mass, Debye-Waller factor, and partial PDOS of the $i$th atom~\cite{RevModPhys.62.1027}. $S(\bm{Q},E)$ is related to $G_i(E)$ by
\begin{equation*}
\label{eq:spdos}
    S(\bm{Q},E)=\sum_{i}\sigma_{i}\frac{\hbar|\bm{Q}|^{2}}{2m_{i}} \textrm{e}^{-2W_{i}} \frac{G_{i}(E)}{E}[1+n(E)].
\end{equation*}
The resulting weighted PDOS for both samples at $T=7$~K are shown in Fig.~\ref{fig:pdos2}(a). It is clear that there are some differences for the two samples, especially at energies below 3~meV. For the superconducting sample, $G(E)$ appears to be significantly higher than that of the nonsuperconducting sample at low energies.

To provide context for these experimental observations, we have performed phonon and electron-phonon coupling calculations in the framework of the density functional perturbation theory~\cite{DFPT}. First-principles calculations were performed using the Quantum-Espresso code~\cite{QE}. Ultrasoft pseudopotentials with Perdew-Zunger type~\cite{PZ-LDA}
local density approximation exchange correlations were generated by the virtual crystal approximation method~\cite{VCA}. Energy cutoffs of 100 Ry for wave functions and 600 Ry for the charge density were employed. The self-consistent calculations were performed over a $12 \times 12 \times 12$ $k$-point grid. A denser $28 \times 28 \times 28$ $k$-grid was used for evaluating an accurate electron-phonon interaction matrix. Dynamical matrices and the electron-phonon coupling were calculated on a $4 \times 4 \times 4$ $\bm{Q}$-point mesh.

The calculated PDOS for both the parent and doped samples are shown in Fig.~\ref{fig:pdos2}(b). Overall, the peaks found in the calculations give a good description (after a slight energy scaling) of many features of the experimental results. The calculated isotropic electron-phonon coupling constant $\lambda$ for (Pb$_{0.5}$Sn$_{0.5}$)$_{0.7}$In$_{0.3}$Te is about 1.4. We use the Allen-Dynes modification of the McMillan equation~\cite{Allen1975},  $T_c=\omega_{\rm{ln}}/1.2\exp[-1.04(1+\lambda)/\lambda-\mu^{\ast}(1+0.62\lambda)]$, where $\omega_{\rm{ln}}$ is the prefactor, and $\mu^{\ast}$ is the Coulomb pseudopotential, with a typical value of 0.13. The $T_c$ is estimated to be around 7.6~K, indicating that electron-phonon coupling is, in principle, compatible with the actual $T_c$.

But is there anything wrong with this story? To begin with, the calculated PDOS completely miss the experimental peaks below 3~meV. To appreciate this discrepancy, we first need to acknowledge that the peaks in the PDOS are van Hove singularities that occur when the derivative of a phonon frequency with respect to the wave vector goes to zero. This typically occurs at a zone boundary, which then involves a substantial volume fraction of the Brillouin zone. The lowest-energy van Hove singularity tends to correspond to the zone-boundary energies of transverse acoustic (TA) phonons. In PbTe (SnTe), the TA mode at the $(1,1,0)$ zone-boundary point has an energy of about 4~meV (6~meV) at $T=50$~K \cite{cowl69,li14}. The average of these values, 5~meV, corresponds to the first big peak in Fig.~\ref{fig:pdos2}(b) and a large peak in Fig.~\ref{fig:pdos2}(a). Our mystery peaks are far below this.

The ferroelectric instabilities in both SnTe and SrTiO$_3$ involve a soft, zone-center, transverse optical phonon \cite{pawl66,onei17,cowl62}.  Such a mode would be unlikely to yield a significant peak in the PDOS, because the zone-center involves a small fraction of the Brillouin-zone volume. In our case, however, the mixture of Pb and Sn (and In) on the same face-centered-cubic sublattice breaks the symmetry of the rocksalt crystal structure, and may allow a mixing of zone-center and zone-boundary phonons. Furthermore, the atomic disorder might result in some localized phonon modes, similar to ones discussed in a model calculation for PbTe \cite{zhan11}.

\begin{figure}[b]
\centering
\includegraphics[width=0.95\linewidth]{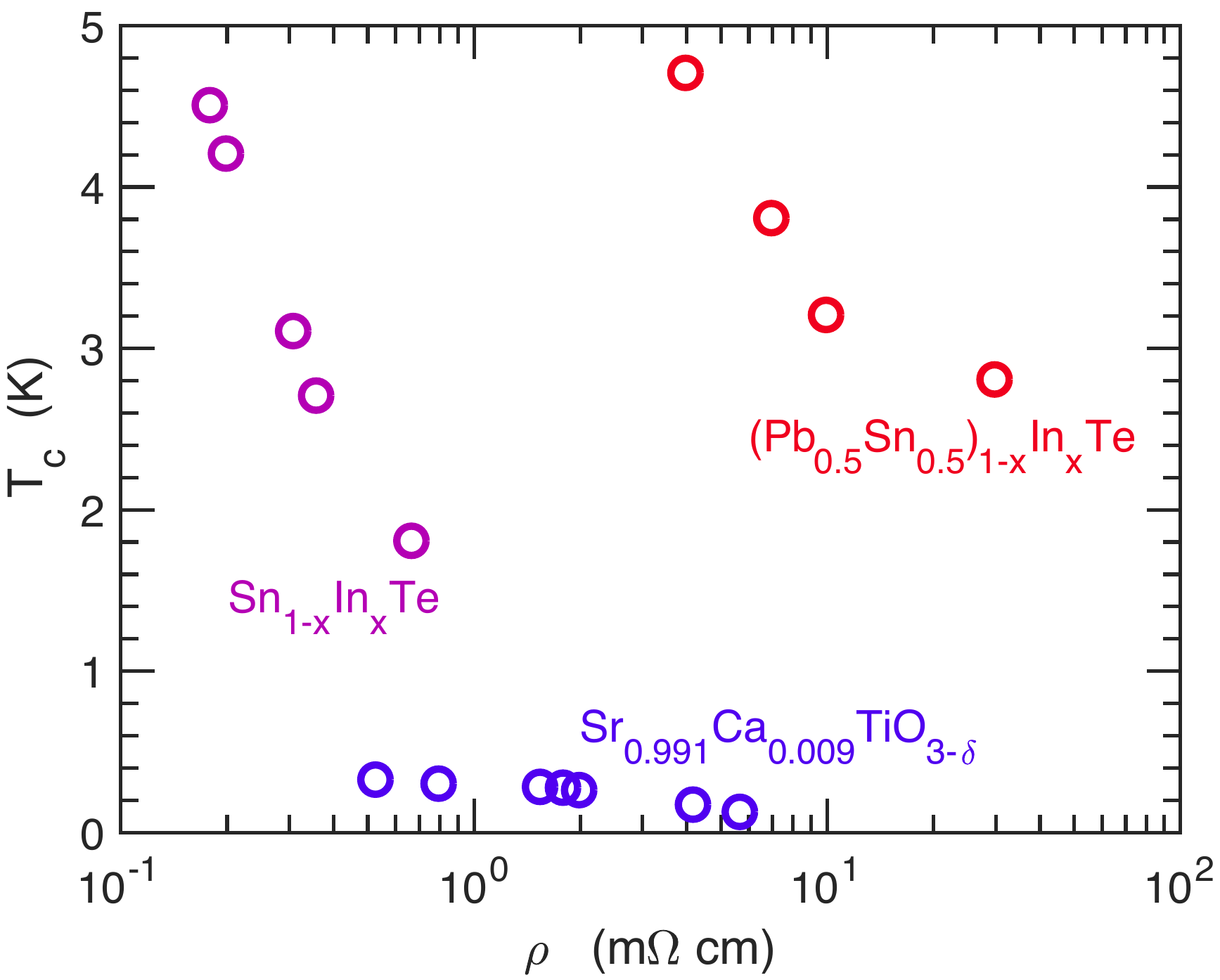}
\caption{\label{fig:rho} Comparison of superconducting $T_c$ vs normal-state resistivity, measured just above $T_c$, for (Pb$_{0.5}$Sn$_{0.5}$)$_{1-x}$In$_x$Te \cite{zhon14},  Sn$_{1-x}$In$_x$Te \cite{PhysRevB.88.020505}, and  Sr$_{0.991}$Ca$_{0.009}$TiO$_{3-\delta}$ \cite{risc17}.}
\end{figure}

Now how do we interpret the change across $T_c$ of the low-energy modes? For a conventional BCS superconductor Nb$_3$Sb, Axe and Shriane showed that the linewidth for the phonons with energies lower than $2\Delta$ became sharper below \tc~\cite{PhysRevLett.30.214,PhysRevB.8.1965}. Several more reports appeared subsequently following these observations~\cite{Shirane19731893,PhysRevB.12.4899,PhysRevLett.77.4628,PhysRevLett.101.237002,PhysRevLett.87.017005,doi:10.1143/JPSJ.70.1480,PhysRevB.55.R8678}.
These results were successfully explained by Allen and co-workers by considering the coupling between conduction electrons and phonons~\cite{Allen1974937,PhysRevB.56.5552}. In particular, the electrons are gapped below $2\Delta(T)$. Thus, damping to the phonons caused by the electron-phonon interaction vanishes for phonons with energies less than 2$\Delta(T)$, resulting in an increase of the phonon lifetime~\cite{Allen1974937,PhysRevB.56.5552}. In this case, the total spectral weight of the phonon scattering should be conserved. Our results in Fig.~\ref{fig:pdos} showing that the PDOS increases for $E\le2\Delta$ at low temperatures but no difference is observed for $E>2\Delta$ are totally unexpected. Nevertheless, this anomalous phonon behavior does indicate the significant electron-phonon coupling known to be present in this class of compounds \cite{luco73,litt79,li14}.

In addition, we have In dopants; in the related compound Sn$_{1-x}$In$_x$Te, each In ion appears to donate one electron, but these electrons only gradually delocalize and become mobile with increasing $x$ \cite{zhan18}.  In the present case, the added carriers must be involved in tuning the superconductivity; however, the normal-state resistivity is very large, and has a nonmetallic temperature dependence \cite{zhon14,crystals7_55}. To illustrate this, in Fig.~\ref{fig:rho} we plot $T_c$ vs $\rho$ (measured just above $T_c$) in PSIT, and compare it with results for Sn$_{1-x}$In$_x$Te \cite{PhysRevB.88.020505,PhysRevB.93.024520} and Sr$_{0.991}$Ca$_{0.009}$TiO$_{3-\delta}$ (STO) \cite{risc17}. The carrier density is at least an order of magnitude greater in PSIT than in STO, but the resistivity is much larger in PSIT. We also note that in Sn$_{1-x}$Ag$_x$Te, although the resistivity just above $T_c$ is lower than 0.1~m$\Omega$~cm, the $T_c$ is only about 2~K \cite{doi:10.7566/JPSJ.85.053702}.

When we cool PSIT below $T_c$, the presumed quasiparticles that participate in the pairing are expected to decouple from the phonons.  If these quasiparticles help to screen the atomic disorder, then decoupling might effectively increase the weight of localized phonons at energies below $2\Delta$.  We are not aware of any previous case where such behavior has been observed.  This unusual behavior certainly merits further experimental and theoretical investigations. Of particular interest to understand is whether well-defined quasiparticles exist in the normal state, as they are the starting point for BCS theory \cite{emer95b}.

In conclusion, we have presented INS results on the phonon spectra of PSIT, with $x=0$ and 0.3. We observe anomalous low-energy features in the PDOS and a substantial enhancement of the phonon intensities below $T_c$, only in the superconducting sample, for phonons with energies below almost exactly twice the zero-temperature superconducting gap. While calculations indicate an electron-phonon coupling strength fully compatible with the $T_c$, they do not explain the anomalous responses found in (Pb$_{0.5}$Sn$_{0.5}$)$_{0.7}$In$_{0.3}$Te. Further study is necessary to understand whether the unusual behavior we have identified is a help or hindrance to the superconducting order.

We thank Xiangang Wan and Phil Allen for stimulating discussions.
Work at Nanjing University was supported by the MOST of China (Grants No. 2016YFA0300404 and No. 2015CB921202), and NSFC (Grants No. 11674157, No. 51372112 and No. 11574133), NSF Jiangsu province (Grant No. BK20150012), the Fundamental Research Funds for the Central Universities (Grant No.  020414380105), and Special Program for Applied Research on Super Computation of the NSFC-Guangdong Joint Fund (the second phase). We also acknowledge the technical support from the HPCC of Nanjing University and ``Tianhe-2" at NSCC-Guangzhou, where the calculations were performed. Work at BNL was supported by the Office of Basic Energy Sciences, U.S. Department of Energy under Contract No. DE-SC0012704. R.D.Z. was supported by the Center for Emergent Superconductivity, an Energy Frontier Research Center, headquartered at BNL, funded by US Department of Energy, under Contract No. DE-2009-BNL-PM015. Research conducted at ORNL's SNS was sponsored by the Scientific User Facilities Division, Office of Basic Energy Sciences, U.S. Department of Energy.



\end{document}